\documentclass[twocolumn,useAMS]{mn2e}
\usepackage{graphicx}
\usepackage{amsmath, amssymb}

\def\etal{{et al.}\thinspace}

\begin{document}

\title[Low frequency m=1 normal mode oscillations]
{Low frequency m=1 normal mode oscillations of a self-gravitating disc}
\author[K. Saha]
       {Kanak Saha \thanks{E-mail : kanak@physics.iisc.ernet.in}\\
   Department of Physics, Indian Institute of Science, Bangalore 560012, India\\} 

\maketitle

\begin{abstract}
A continuous system such as a galactic disc is shown to be well approximated by an N-ring differentially
rotating self-gravitating system. Lowest order (m=1) non-axisymmetric features such as lopsidedness and warps are global in nature and quite common in the discs of spiral galaxies. Apparently these two features of
the galactic discs have been treated like two completely disjoint
phenomena. The present analysis based on an eigenvalue approach brings
out clearly that these two features are fundamentally similar in
nature and they are shown to be very Low frequency Normal Mode (LNM) oscillations manifested in different symmetry planes of the galactic disc. Our analysis also show that these features are actually
long-lived oscillating pattern of the N-ring self-gravitating system.


\end{abstract}

\section{Introduction}

A typical N-ring differentially rotating self-gravitating system in astrophysical context is the stellar galactic disc or the gaseous disc. Disk galaxies often harbour a lot of 'beautiful and conspicuous' mostly non-axisymmetric patterns in the form of the spectacular spiral structures, bars, rings, warps, lopsidedness, scalloping. All these patterns have been a mystery for a long time and they continue to be even today! It is not obvious why these patterns are formed, how they are formed and how they are maintained in the discs of spiral galaxies? Depending on the orientations of these disc galaxies in the sky with respect to the line-of-sight (LOS) various patterns become available for direct astronomical observations. In cylindrical polar coordinate system (R, $\varphi$, z), suitable for the disc geometry, perturbations describing many such patterns are of the type $\sim$ $e^{i m \varphi}$, $\varphi$ being the azimuthal coordinate and m is the azimuthal wavenumber. Lowest order modes especially the m=1 are of great interest in astrophysics. For perturbations normal to the equatorial plane of the disc, m=1 represents a very commonly observed configuration of the disc known as warps. Warps are systematic bending (Lynden-Bell 1965; Hunter \& Toomre 1969) of the galactic equatorial plane on either side of the galactic centre in the form of a characteristic anti-symmetric integral sign shape (when viewed with an inclination angle $i \sim 90\,^{\circ}$ between the LOS and the minor axis of the galactic disc). Whereas in the equatorial plane of the disc ($i \sim 0\,^{\circ}$) such m=1 perturbations represent another very commonly observed phenomenon called the lopsidedness. Lopsidedness is referred to as the large scale asymmetry in the observed distribution of the stellar light (Block et al. 1994; Rix \& Zaritsky 1995) as well as in the distribution of atomic hydrogen gas (Baldwin et al. 1980; Richter \& Sancisi 1994; Hynes et al. 1998). Statistical analysis of various observations coming from a broad range of the electromagnetic spectrum (visible - IR - radio) have established the fact that warps and lopsidedness are ubiquitous in disc galaxies. They are almost proving to be an universal pattern in spiral galaxies. Being so abundant in nature, such astounding patterns put forward a question if they are primordial or repeatedly created along their evolutionary path. Since the strong differential shear in the galactic disc would wash away any such coherent pattern leading to a phase-mixed equilibrium within a billion year time scale, the survival of these patterns intact over 10 billion years is a more serious issue. Despite being so common there is still no fully convincing and unquestionable mechanism for the origin of these patterns in galactic discs. Warps and lopsidedness being an outer disc phenomena can be used as a potential probe for investigating the nature of gravity in the outskirts of the galactic disc, the structure of the dark matter halos. These patterns can also put valuable constraints on the galaxy evolution and galaxy formation processes (Ostriker \& Binney 1989; Binney 1992; Combes 2004). So it is important to have a deeper understanding about the nature of these patterns in the galactic disc. Even though warps and lopsidedness are two completely diverse phenomena, normally observed in two different class of objects face-on ($i\sim 0\,^{\circ}$) and edge-on ($i\sim 90\,^{\circ}$) galaxies, there exists an intrinsic similarity between the two observed patterns. The aim of the present letter is to show that these two different phenomena are fundamentally similar in nature. We attempt to present an analysis based on an eigenvalue approach to bring out the fact that these phenomena are basically discrete low frequency normal mode oscillations of the galactic discs.

\noindent We consider a razor thin, initially axisymmetric cold self-gravitating disc having an exponential distribution of mass surface density, $\Sigma^{0}(R)=\Sigma_{0}e^{-R/R_d}$, with a scale length $R_d$ and central surface density $\Sigma_{0}$. The unit of frequency is $\sqrt{\pi G \Sigma_{0}/R_d}$ $\sim 53.0$ kms$^{-1}$kpc$^{-1}$ for a typical disc like that of Milky Way with $\Sigma_{0}=640.9$ M$_{\odot}$pc$^{-2}$ and R$_d$=3.2 kpc. Note that these parameters are chosen to represent a typical disc model with disc mass, M$_d \sim$ 4$\times 10^{10}$ M$_{\odot}$. The disc is differentially rotating with angular speed $\Omega(R)$ so that the unperturbed velocity field is given by $\vec{v} = (0,R\Omega,0)$. In the linear approximation for a razor thin disc (z/R $<<$ 1) the vertical dynamics is almost completely decoupled from the planar one and as a consequence we are justified in treating lopsidedness and warp as two independent phenomena in the disc.
  
\section{Lopsidedness in a cold self-gravitating disc}
We formulate here the dynamical equations governing the global lopsided mode in the razor thin galactic disc by assuming hydrostatic equilibrium along the vertical direction. We write the velocity field, surface density and potential of the perturbed disc as:
 $v_{R}=v_{R}^{\prime}$, $v_{\varphi}= R\Omega + v_{\varphi}^{\prime}$, $\Sigma = \Sigma^{0} + \Sigma^{\prime}$, $\Phi = \Phi^{0} + \Phi^{\prime}$. 
\noindent Then the linearized Euler and continuity equations can be written as:
\begin{eqnarray}
&& \frac{Dv_{R}^{\prime}}{Dt} - 2\Omega v_{\varphi}^{\prime} \:=\: -\frac{\partial{\Phi^{\prime}}}{\partial R}\\
&& \frac{Dv_{\varphi}^{\prime}}{Dt} + \frac{\kappa^2}{2\Omega} v_{R}^{\prime} \:=\: -{\frac{1}{R}}\frac{\partial{\Phi^{\prime}}}{\partial \varphi}\\
&& \frac{D\Sigma^{\prime}}{Dt} + {\frac{1}{R}}\frac{\partial(R\Sigma^{0}v_{R}^{\prime})}{\partial R} \:=\: -{\frac{\Sigma^{0}}{R}}\frac{\partial{v_{\varphi}^{\prime}}}{\partial \varphi} 
\end{eqnarray}
 
\noindent In the above equations ${D}/{Dt}\equiv {\partial}/{\partial t} + \Omega {\partial}/{\partial \varphi}$ and $\kappa$ is the radial epicyclic frequency in the disc.

We consider all the perturbed variables as $X^{\prime}(R,\varphi,t) = X^{\prime}(R)e^{i(\varphi - \omega t)}$ about the equilibrium state. 
\noindent Stability of the mode depends on the sign of the imaginary part of $\omega$. Lopsided modes in the disc become unstable when Im($\omega$) $> 0$. They are stable decaying modes when Im($\omega$)$ < 0$. Modes with Im($\omega$) $= 0$ are in general stationary van Kampen modes; outside the continuum, these modes are pure normal mode oscillation of the whole disc.

Substituting the above form of the perturbed variables into eqs.(1-3) and combining the resulting velocity field and the perturbed continuity equation in the slow mode limit ($\omega << \Omega$), we get an implicit relation between the perturbing potential and the perturbed surface density of the disc. 
\noindent Now we require the perturbed surface density($\Sigma^{\prime}$) and the potential ($\Phi^{\prime}$) to be connected through the Poisson equation in order to produce a self-consistent solution of the problem. This we achieve using the integral form of the Poisson equation for the perturbed disc under the imposed $m=1$ lopsided mode:

\begin{equation}
\Phi^{\prime}(R) = - G \int_{0}^{\infty}{d R^{\prime} R^{\prime} \mathcal{H}_{lop}(R,R^{\prime}) \Sigma^{\prime}(R^{\prime})}
\end{equation}

\noindent Where the kernel in eq.(4) is given by
\begin{equation}
\mathcal{H}_{lop}(R,R^{\prime}) = \int_{0}^{2\pi} \frac{\cos{\alpha}  d\alpha}{[R^2 + {R^{\prime}}^2 - 2 R R^{\prime}\cos{\alpha} + b^2]^{\frac{1}{2}}}  - \pi \frac{R}{{R^{\prime}}^2}
\end{equation}
 
\noindent The kernel represents the softened self-gravity of the perturbation, $b$ being the softening parameter. This allows us to perform the numerical integration over the disc by removing the singularity at $R=R^{\prime}$. The second indirect term in the kernel arises due to the $m=1$ lopsided mass distribution about the geometrical centre of the disc. The indirect term plays a crucial role in making the disc susceptible to the lopsided instability. 

Using the perturbed potential (eq.[4]) and performing straightforward mathematical manipulations, we arrive at the following integral equation describing the global dynamical behaviour of the m=1 lopsided mode under the collective effect of the softened self-gravity:

\begin{equation}
{\omega^2}\Sigma^{\prime} + \mathfrak{D}_{lop}(\Sigma^{\prime}) \omega + \mathcal{S}_{lop}(\Sigma^{\prime}) = 0 
\end{equation}

\noindent  Eq.(6) can be solved by recasting it into a matrix-eigenvalue problem. By discretizing on a uniform grid with $N$ radial points in the disc resembling a concentric N-ring system we can write eq.(6) in a compact form (for details see Saha et al. 2007):

\begin{equation}
 \left[{\omega^2}I +\omega \mathfrak{D}_{lop} + \mathcal{S}_{lop}\right] \Sigma_{lop} = 0
\end{equation}

Where $\Sigma_{lop}$ is the eigenvector corresponding to the eigenvalue $\omega$. I, $\mathfrak{D}_{lop}$, and $\mathcal{S}_{lop}$ are the three $N \times N$ real square matrices. We call I the unit mass matrix,
$\mathfrak{D}_{lop}$ as the general damping matrix and $\mathcal{S}_{lop}$ as the stiffness matrix for the galactic disc decomposed into an N-ring system. 

\section{Warping of the cold self-gravitating disc}

The dynamical equation of a small bending of the disc perpendicular to 
its unperturbed plane($z=0$) is given by

\begin{equation}
\frac{D^{2}z_{1}}{D t^2} = {{\mathcal F}_{self}} +{{\mathcal F}_{halo}}
\end{equation}
 
\noindent where $D/Dt$ is defined in the previous section and the small bending is described by a single function ${z_{1}}(R,\varphi,t)$. ${{\mathcal F}_{halo}}$ is the vertical restoring force near the disc equatorial plane ($z=0$) due to a screened isothermal dark matter halo (de Zeeuw \& Pfenniger 1988).

\begin{equation}
{\mathcal F}_{halo}(R,\varphi,t)  =  -{\nu_{h}}^2 (R)z_{1}(R,\varphi,t)
\end{equation}

\noindent where $\nu_{h}$ is the local vertical frequency due to the dark matter halo. And ${\mathcal F}_{self}$ is the vertical self-force due to the bent disc itself:

\begin{eqnarray}
\lefteqn{{{\mathcal F}_{self}(R,\varphi,t)} = -G{\int_{0}^{\infty}}{\Sigma^{0}(R^{\prime})R^{\prime} dR^{\prime}} }  \nonumber \\
& & \times{ {\int_{0}^{2\pi}} \frac{[z_{1}(R,\varphi,t) -z_{1}(R^{\prime},\varphi^{\prime},t)]}{[R^2 + {R^{\prime}}^2 -2 R R^{\prime}\cos(\varphi -\varphi^{\prime}) + {b}^2]^{\frac{3}{2}}} d\varphi^{\prime} } \nonumber \: \: \: \: \: \: \: \: \: (10)
\end{eqnarray}

\noindent In the above equations $b$ is again the softening parameter. 

\noindent  $z_{1}(R,\varphi,t) = \Re\{ h_{w}(R) e^{i(\varphi - \omega t )}\}$ describes the bending mode of the disc. 
\noindent Substituting $z_{1}$ into eq.[8] and discretizing it on a uniform grid with N radial points in the disc we have:

$$\left[ {{\omega}^2}I + {\omega}D_{w} + S_{w} \right ]h_{w}  =  0 \eqno(11) $$

\noindent Where I, $D_{w}$, and $S_{w}$ are the three N$\times$N real square matrices. Once again, I is the unit mass matrix, $D_{w}$ is the damping matrix and $S_{w}$ is the stiffness matrix (for details see Saha \& Jog 2006). $h_{w}$ is the eigenvector corresponding to the eigenvalue $\omega$. 

\noindent Both eq.[7] and eq.[11] represent a class of nonlinear eigenvalue problem (here quadratic eigenvalue problem, hence QEP) to describe the global behaviour of the m=1 modes in a self-gravitating disc.

\begin{figure}
{\rotatebox{270}{\resizebox{4.5cm}{5.5cm}{\includegraphics{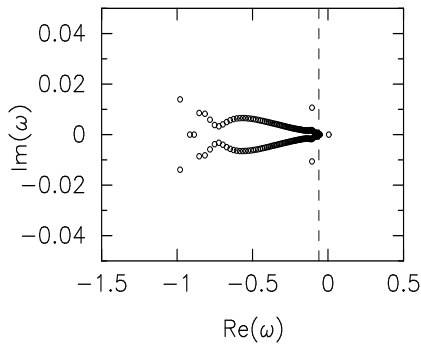}}}}

\caption{
Complete eigen spectrum of a cold lopsided disc. Dashed line denotes lower boundary of the continuum ($\Omega -\kappa$). The isolated point corresponding to $Im(\omega)=0$ on the right side of the dashed line lies in the P-gap and represent global discrete lopsided mode. }
 \end{figure}

\begin{figure}
{\rotatebox{270}{\resizebox{4.0cm}{5.0cm}{\includegraphics{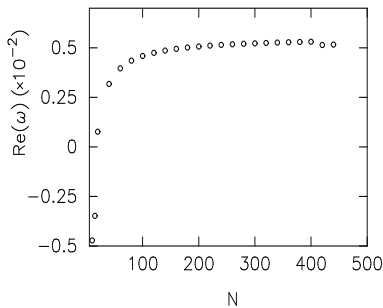}}}}

\caption{
The pattern speed of a cold lopsided disc. For a disc like that of our galaxy, this pattern speed is $\sim 0.27$ km s$^{-1}$kpc$^{-1}$.}
\end{figure}

\section{Complete eigen-spectrum of the N-ring system}
We treat  the galactic disc as a discrete N-ring system for studying its two very important global patterns. Both eqs.[7,11] are finite N-dimensional QEP representing a class of nonlinear eigenvalue problem. We solve such QEPs by linearizing it into a 2N$\times$2N eigensystem of the type $A y=\lambda y$. In a typical scenario, the disc particles oscillate vertically with frequency $\nu$ and radially with epicyclic frequency $\kappa$ while they circulates around the galactic centre with circular frequency $\Omega$. A local WKB analysis of the m=1 type perturbations in the continuous disc tells us that in the absence of any self-gravity any vertical m=1 disturbance (warping mode) would precess in the disc with two frequencies $\Omega \pm \nu$. Similarly, any m=1 disturbance in the equatorial plane (lopsided mode) of the disc without any self-gravity would precess with $\Omega \pm \kappa$. The '+' sign denotes a fast mode while '-' is for the slow mode. In galactic dynamics, the slow modes are of particular interest because they can survive longer compare to the fast mode against the differential shear in the disc. In normal galactic disc since $ \nu, \kappa > \Omega $, the slow mode precession frequencies are always negative i.e. they are the retrograde modes in the disc.   
Naturally, according to WKB analysis there exists a principal gap (hence P-gap) of length $2\nu$ in the allowed precession frequencies for the warps and $2\kappa$ in the case of lopsidedness in the discs of spiral galaxies. In absence of self-gravity, there exists no mode in the P-gap which is acting like a forbidden gap in the semiconductors! The scenario changes qualitatively when there is self-gravity associated with each of these patterns in the global sense.

\begin{figure}
{\rotatebox{270}{\resizebox{4.5cm}{5.5cm}{\includegraphics{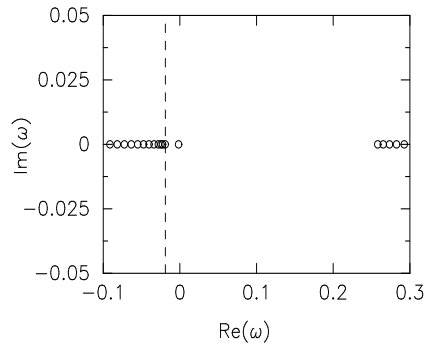}}}}

\caption{
Complete eigen spectrum of a cold warped disc. Dashed line denotes lower boundary of the continuum ($\Omega -\nu$). The isolated point corresponding to $Im(\omega)=0$ on the right side of the dashed line is in the P-gap and represents the discrete global warping mode of the disc.}
\end{figure}

On solving the N-dimensional QEP (eq.[7]) for the lopsided perturbations, we obtain 2N eigenvalues and these eigenvalues are plotted in Fig. 1. We primarily concentrate on those slow modes for which  Re($\omega$) $\sim$ 0. The eigen spectrum shows that there exists a distinct (clearly separated from the rest of the points in the Argand diagram) point in the P-gap, close to the boundary of the lower continuum ($\Omega - \kappa$), corresponding to Re($\omega$) $\sim$ 0 and Im($\omega$)=0. The mode corresponding to this point is an almost non-rotating stationary global lopsided pattern of the disc. Being isolated in the Argand diagram, this mode resembles like an excited normal mode of the disc. Stationary normal mode oscillations of spherical stellar system has been studied in detail by many authors (Miller \& Smith 1995; Vandervoort 2003) previously. The existence of a normal mode of oscillation outside the continuum was shown by Mathur (1990) for 1D and 3D gravitating simplistic system. The presence of such a distinct normal mode in our eigen solution for a disc-like 2D self-gravitating system confirms the previous trend found in theoretical studies namely self-gravitating systems are capable of supporting normal modes of oscillation. In Fig. 2, we show that the position of this mode in the complex plane is not sensitive to the number of rings used to represent the disc once N is moderately larger. It shows that the pattern speed (=$Re(\omega)$) of the lopsided mode stays almost constant as we increase the number of rings. Thus it would not be unfair to conclude that for moderately large N, an N-ring discrete system is quite capable of representing the basic physics of a continuous system like the galactic disc. 
Note that the above two figures are derived for a purely galactic disc without any dark matter halo. However, including a rigid, non-responsive screened isothermal dark matter halo (with M$_h \sim 6.3$ M$_d$ within 10 R$_d$ and producing a reasonable flat rotation curve), we find that the spectrum of the lopsided mode does not change qualitatively. As the dark matter halo mass increases, the eigen frequency of the isolated discrete normal mode (lying in the P-gap) shifts towards the lower boundary of the continuum and in the limit of dark matter too dominating over the disc, the discrete normal mode ceases to exist. But in reality, the dark matter halo may not be non-responsive and such live halo might enhance the self-gravity of the perturbation which in turn would help sustaining the discrete normal mode oscillation lying in the P-gap (which we would like to study in future).
The spectrum of lopsided N-ring system is quite interesting in the sense that it supports a family of unstable modes including a very few van Kampen type stationary modes in the continuum region. We have checked that the behaviour of the eigen spectrum in the continuum region does not change with the inclusion of a dark matter halo.  

The complete spectrum of the warped disc is shown in Fig. 3. 
In deriving the eigen spectrum for the warp, we have included a dark matter halo of mass M$_h$ $\sim 6.3 \times$ M$_d$ calculated within 10 R$_d$. The total mass of the galaxy comes out to be $\sim 3.0 \times 10^{11}$ M$_{\odot}$ for the assumed disk-halo model which produces a flat rotation curve.
The real part of the eigen values represent the pattern speed of the warping mode. The N-ring system produces 2N such pattern speeds. All the eigen modes are purely oscillatory. The spectrum contains a P-gap  of length exactly the same size as twice the local vertical frequency (2$\nu$). Fig. 3 clearly shows the lower and upper boundary of the continuum. And there exists one very distinct mode in the P-gap resembling an integral sign  warp as seen in the observations.
In the present study, we particularly emphasize on the behaviour of the N-ring system representing the actual galactic disc which is basically a continuous system. However, even the galactic disc could be grainy if we look at it beyond the epicyclic scale which is typically much less than the scale length of the underlying mass distribution. So unless we go down beyond the epicyclic scale the galactic disc more or less can be considered as a continuous system. We try to represent the behaviour of such a continuous system by an N-ring differentially rotating self-gravitating system.
 
\begin{figure}
{\rotatebox{270}{\resizebox{4.0cm}{5.0cm}{\includegraphics{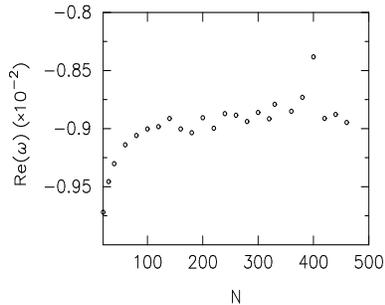}}}}

\caption{
Pattern speed of the warped disc. For a typical disc like that of Milky Way, the pattern frequency $\sim 0.47$ km s$^{-1}$kpc$^{-1}$.}
\end{figure}

In Fig. 4 we show how the eigenvalue of the discrete mode in the P-gap changes as we increase the number of rings (N) to represent the continuous disc. In the limit of a large N (in statistical sense) the pattern speed of the discrete mode does not vary appreciably. The position of the eigenvalue remains more or less same in the Argand diagram. This directly confirms the fact that there exists atleast one such discrete normal mode of oscillation of the N-ring system and it stands out in glorious fashion amongst the sea of continuum modes to represent a warp of the disc. 

\begin{figure}
{\rotatebox{270}{\resizebox{4.0cm}{5.0cm}{\includegraphics{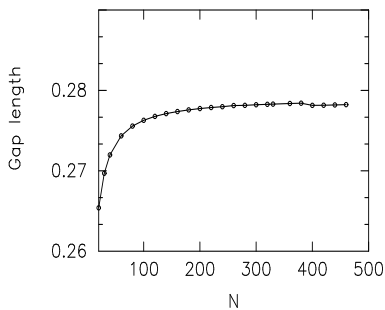}}}}

\caption{ Size of the P-gap in the eigenvalue spectrum of the warped disc.}
\end{figure}
\noindent In Fig. 5, we calculate the length of the P-gap as seen in Fig. 3. The dots are the one from the eigenvalue analysis. As N increases beyond about 100, the gap length evaluated from the eigenvalue analysis exactly matches the analytical value $\sim 0.276$ calculated for the assumed disc-halo model. This shows a beautiful match between the two. 
\section{Discussion and Conclusions}

From the present analysis, it is clear that global self-gravity plays a crucial role in governing the behaviour of warps and lopsidedness in galactic discs. Self-gravity also plays a key role in another interesting galactic system known as polar rings (PRs) found around S0 galaxies. PRs are rotating self-gravitating system and differential precession should destroy the flatness of the PRs (Sparke 1986; Anraboldi \& Sparke 1994) within a billion year time scale. The fact that we see such beautiful polar rings having almost the same inclination about the equatorial plane of the host galaxy, implies that self-gravity must be playing a major role in maintaining such feature. This very basic idea is common in the present analysis of warps, lopsidedness and that of polar rings by Sparke (1986). So implementing this idea, one can in principle construct an integro-differential equation for the inclinations of the polar rings/disk and seek a normal mode like solution. Basically, understanding of these phenomena then reduces to answering the following question: Can we find a bound state (here, a discrete normal mode) of the problem in the presence of global self-gravity (here, acting like a potential well)? Remember, in the absence of a self-gravity, these features are just the free plane waves (lying in the continuum) which are going to get wind up due to the strong differential precession. 

One of the possible ways to solve such an integral equation is to convert it into a matrix eigenvalue problem (Kalnajs 1977; Polyachenko 2005). For an infinitely large continuous physical system this matrix eigenvalue problem is also infinite dimensional. In order to numerically solve such an infinite dimensional matrix eigenvalue problem one has to truncate it artificially and make it a finite N-dimensional matrix eigenvalue problem solvable by computer. Such a truncation is meaningful only when the kernel generating the matrix is convergent as N becomes larger i.e. the norm of the kernel remains finite. So the basic idea here is to see if such a finite N-dimensional eigenvalue problem can recover the basic physics of the parent continuous system. Our numerical analysis shows that an N-ring differentially rotating self-gravitating system is an adequate representation of a continuous system like galactic discs. 
Note that this idea of N-ring has also been used previously in the studies of polar rings (Sparke 1986). Apparently, there seems to be a difference between the two analyses. Our N-ring system is a mathematical construction, while the N-rings (each having mass $m_i$) used in Sparke's analysis is physical. In any case, the methods should converge as N becomes very large.
In the linearized version of the two QEPs governing the dynamics of lopsided patterns and warps, the norms of the 2N-dimensional matrix operators are finite and thus are of Hilbert-Schmidt type operators having a discrete spectrum of size 2N. Numerical analysis of the eigen spectra shows that warps and lopsidedness in a cold disc are fundamentally similar in nature. 
Both of them are very slowly (pattern speeds are $\sim$ 2 orders of magnitude less than the oribital frequency) precessing (hence long-lived) normal modes lying in the principal gap of the N-ring self-gravitating disc. Such a unified picture of warps and lopsidedness as a low frequency normal mode oscillation of the N-ring differentially rotating self-gravitating system helps us unraveling the more complicated nature of spiral galaxies. 





\medskip

\noindent {\bf{Acknowledgments}}
The author thanks the anonymous referee for an encouraging and constructive comments on the manuscript. The author acknowledges the current support from Raman Research Institute, India.






\end{document}